\documentclass[global]{svjour}
\pdfoutput=1

\usepackage{latexsym}
\usepackage{graphicx}
\usepackage{dcolumn}

\newcommand{\bra}[1]{\langle #1 |}
\newcommand{\ket}[1]{| #1 \rangle}
\newcommand{\braket}[2]{\langle #1 | #2 \rangle}
\newcommand{\im}{\dot{\iota}\,}
\newcommand{\vecI}{\mathrm{I}}

\begin{document}

\title{Quantum repeater based on cavity-QED evolutions and coherent light}

\author{Denis Gon\c{t}a, Peter van Loock}
%
\institute{Institute of Physics, Johannes Gutenberg University Mainz, \\
             Staudingerweg 7, 55128 Mainz, Germany \\
             \email{dengonta@uni-mainz.de, loock@uni-mainz.de}}
\date{Received: date / Revised version: date}
%
\maketitle

\begin{abstract}
In the framework of cavity QED, we propose a quantum repeater scheme that uses coherent 
light and chains of atoms coupled to optical cavities. In contrast to conventional repeater 
schemes, we avoid the usage of two-qubit quantum logical gates by exploiting solely the 
cavity QED evolution.
In our previous paper [D.~Gonta and P.~van Loock: Phys. Rev. A \textbf{88}, 052308 (2013)], 
we already proposed a quantum repeater in which the entanglement between two neighboring 
repeater nodes was distributed using controlled displacements of input coherent light, 
while the produced low-fidelity entangled pairs were purified using ancillary (four-partite) 
entangled states.
In this paper, the entanglement distribution is realized using a sequence of controlled 
phase shifts and displacements of input coherent light. Compared to previous coherent-state-based
distribution schemes for two-qubit entanglement, the present scheme relies only upon a simple
discrimination of two coherent states with opposite signs, which can be performed in a quantum
mechanically optimal fashion via a beam splitter and two on-off detectors.
The entanglement purification utilizes a scheme that avoids the usage of ancillary entangled 
resources. Our repeater scheme exhibits reasonable fidelities and repeater rates providing an 
attractive platform for long-distance quantum communication. 
\end{abstract}

\section{Introduction}
\label{intro}

In classical data transmission, repeaters are used to amplify data signals 
(bits) when they become weaker during their propagation through the transmission 
channel. In contrast to classical information, the above mechanism is impossible 
to realize when the transmitted data signals are the carriers of quantum 
information (qubits), which cannot be amplified or cloned without destroying 
the encoded information \cite{nat299}, \cite{pla92}. Therefore, the carrier has to 
propagate along the entire length of the transmission channel which, due to 
various losses, leads to an exponentially decreasing probability 
to receive it intact at the end of the channel.

To circumvent this problem, the quantum repeater was proposed in the seminal 
Ref.~\cite{prl81} that encapsulates three building blocks (i) entanglement 
distribution, (ii) entanglement purification, and (iii) entanglement swapping, 
which have to be applied sequentially. Using the first building block, a large 
set of low-fidelity entangled qubit pairs is generated between all repeater nodes, 
which becomes distilled by the second block into a smaller set of high-fidelity 
entangled pairs. Entanglement swapping, finally, combines two purified entangled 
pairs distributed between the neighboring repeater nodes into one entangled pair 
leading to a gradually increasing distance of shared entanglement.

Obviously, quantum repeater schemes are not straightforward. The above
mentioned building blocks, for instance, require a feasible and reliable quantum 
logic, while low-fidelity entangled pairs have to be distributed over reasonably 
long distances. Up to now, only a few schemes, which distribute entanglement over 
the distance of 200 km using fiber-optic \cite{oe17} and 100 km using free-space 
channel \cite{nat488} between the nodes have been experimentally demonstrated. 
Nevertheless, motivated by a rapidly growing experimental and theoretical progress 
in the field of quantum communication, various quantum repeater schemes and single 
building blocks have been proposed \cite{prl96}, \cite{prl101}, \cite{prl105}, 
\cite{rmp83}, \cite{rpr76}, \cite{pra87}.

In our previous paper \cite{pra88a}, we already proposed a dynamical quantum 
repeater scheme in which the entanglement between the two neighboring repeater 
nodes was distributed using controlled displacements of input coherent light, 
while the generated low-fidelity entangled pairs were purified using ancillary 
(four-partite) entangled states. This scheme exploited solely the evolution of 
short chains of atoms coupled to optical cavities located in each repeater node, 
such that any explicit usage of quantum logical gates has been avoided. In the 
present paper, we propose an improved quantum repeater scheme that (as well) 
avoids two-qubit quantum logical gates and where the entanglement is distributed 
using a sequence of controlled phase shifts and displacements of input coherent 
light, while entanglement purification is free of ancillary entangled resources. 
This repeater scheme exhibits reasonable fidelities and repeater rates providing 
an attractive platform for long-distance quantum communication.

In particular, our entanglement distribution becomes very efficient, simple, and
practical, as it only requires discriminating two optical coherent states with 
opposite signs, where each allows for conditionally preparing a certain two-qubit
Bell state over the distance in two neighboring repeater nodes. This way, in the
ideal loss-free case, maximum two-qubit entanglement can be deterministically 
generated with unit fidelity in the limit of large coherent-state amplitudes.
Note that in the present work we focus on a standard (original) quantum repeater 
scheme as described above and based on the use of quantum memories and two-way 
classical communication. Other more recent approaches that make use of quantum 
error correction codes reduce or completely eliminate the need for storing quantum 
information and for employing two-way communication. As a result, extremely high 
rates are achievable in this new generation of repeaters (for a classification into 
three generations of quantum repeaters, see Ref.~\cite{prl112}).

\begin{figure}[!ht]
\begin{center}
\includegraphics[width=0.95\textwidth]{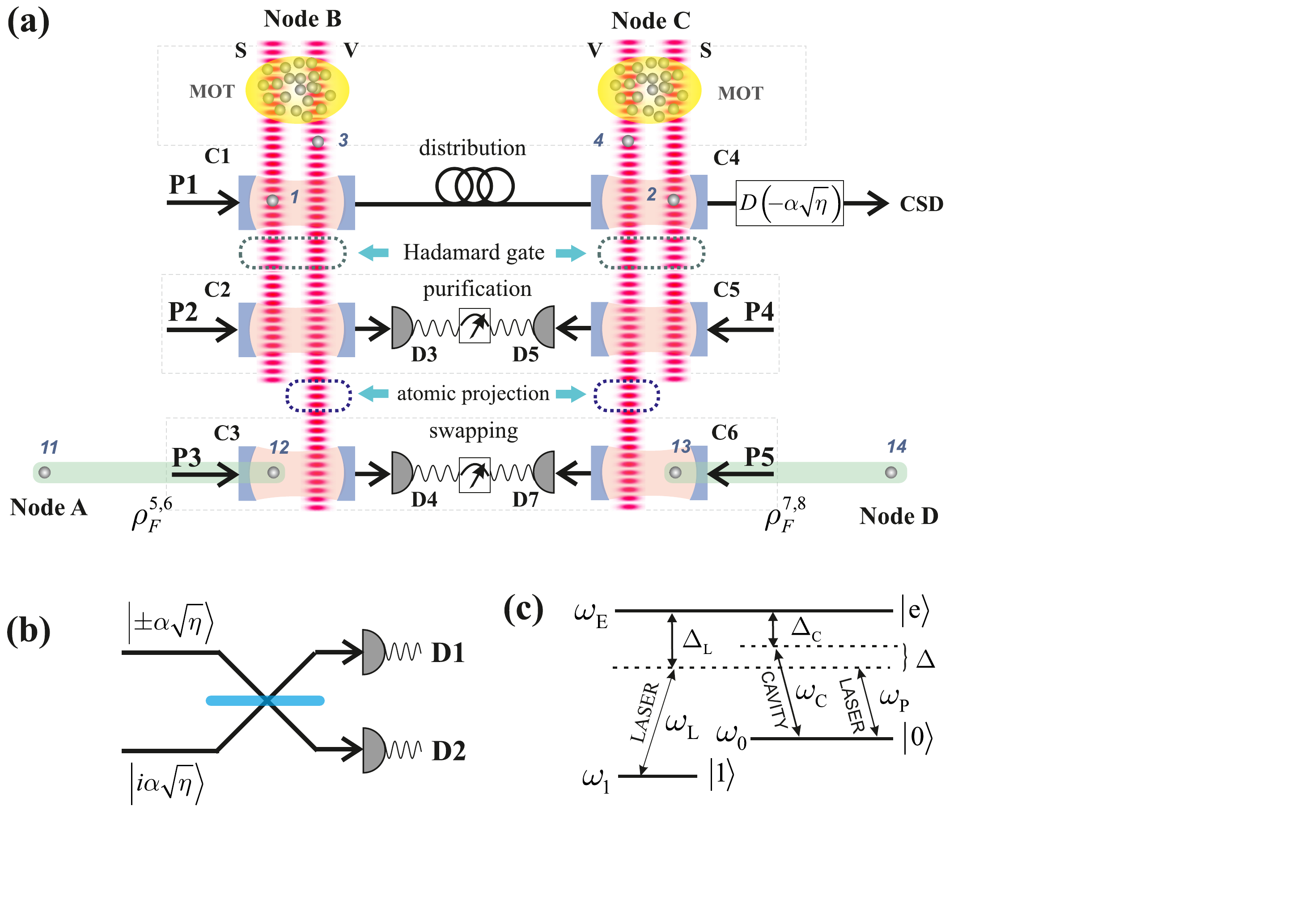} \\
\caption{(Color online) (a) Sketch of experimental setup that realizes the two-node 
repeater scheme. In this setup, atoms in both repeater nodes are synchronously conveyed
with the help of vertical optical lattices starting from the MOTs (oval regions) towards 
the resonators $C_3$ ($C_6$). In the upper region (framed by a dashed rectangle), the 
atomic pairs become (one-by-one) entangled providing the entanglement resources for the 
next framed region, where the entanglement purification takes place. A successful 
purification round leads to an increase of fidelity associated with a stationary atomic 
pair, while the other pairs get projected once they leave the cavities $C_2$ ($C_5$). In 
the final (lower) framed region, the atoms (associated with the purified atomic pair) 
enter the cavities $C_3$ ($C_6$), where they get swapped with atoms (associated with other 
two purified atomic pairs) leading to a triple distance enlargement of shared entanglement. 
(b) The coherent state discrimination (CSD) setup. The coherent states $\ket{\pm \sqrt{\eta} 
\alpha}$ encapsulated in Eq.~(\ref{state0}) are guided into the upper input port of the 
beam-splitter, while the ancilla coherent state $\ket{\im \sqrt{\eta}} \alpha$ into the 
lower input port. (c) Structure of a three-level atom in the $\Lambda$-configuration 
subjected to the cavity and laser fields. See text for description.}
\label{fig1}
\end{center}
\end{figure}

The paper is organized as follows. In the next section, we describe in detail our 
practical quantum repeater scheme. We introduce and discuss the entanglement 
distribution, purification, and swapping protocols in Secs.~II.A, II.B, and II.C, 
respectively. In Sec.~II.D, we discuss a few relevant issues related to 
the implementation of our repeater scheme, while a brief analysis of final 
fidelities and repeater rates is given in Sec.~III along with the summary and
outlook.

\section{Quantum repeater platform}
\label{sec:1}

The main physical resources of our repeater scheme are (i) three-level atoms, (ii) 
high-finesse optical cavities, (iii) continuous and pulsed laser beams, (iv) balanced 
beam splitters, and (v) photon detectors. In Fig.~\ref{fig1}(a), we display schematic 
view of our experimental setup that encapsulates two neighboring repeater nodes (B and 
C) and includes the entanglement distribution, purification, and swapping protocols in
a single setup.

In this setup, each repeater node encapsulates single-mode cavities $C_1$, $C_2$, and 
$C_3$ ($C_4$, $C_5$, and $C_6$), single atoms conveyed along the setup with the help of 
two vertical optical lattices, two sources of weak coherent-state pulses $P_2$ and $P_3$
($P_4$ and $P_5$), detectors $D_3$ and $D_4$ ($D_5$ and $D_6$) connected to the neighboring
node via a classical communication channel, and a magneto-optical trap (MOT) that provides 
atomic chains to be conveyed. The alignment of vertical lattices is such that the conveyed 
atoms cross cavities at their anti-nodes ensuring, therefore, a strong atom-cavity coupling 
regime. In addition, node B includes a source of weak coherent-state pulses $P_1$, while 
node C includes a coherent state discrimination (CSD) setup displayed in Fig.~\ref{fig1}(b). 
Finally, each repeater node encapsulates two optical lattices denoted as S and V (conveyors), 
where the atoms inserted into the V-conveyor are transported with a constant velocity through 
all three cavities, while the atoms in the S-conveyor can be slowed down or accelerated by 
demand.

For convenience, our setup is divided in three (framed by dashed rectangles) parts 
corresponding to three building blocks of a quantum repeater as mentioned in the introduction. 
Below, we clarify each part of our setup and explain all the manipulations and elements.

\begin{figure}[!ht]
\begin{center}
\includegraphics[width=0.95\textwidth]{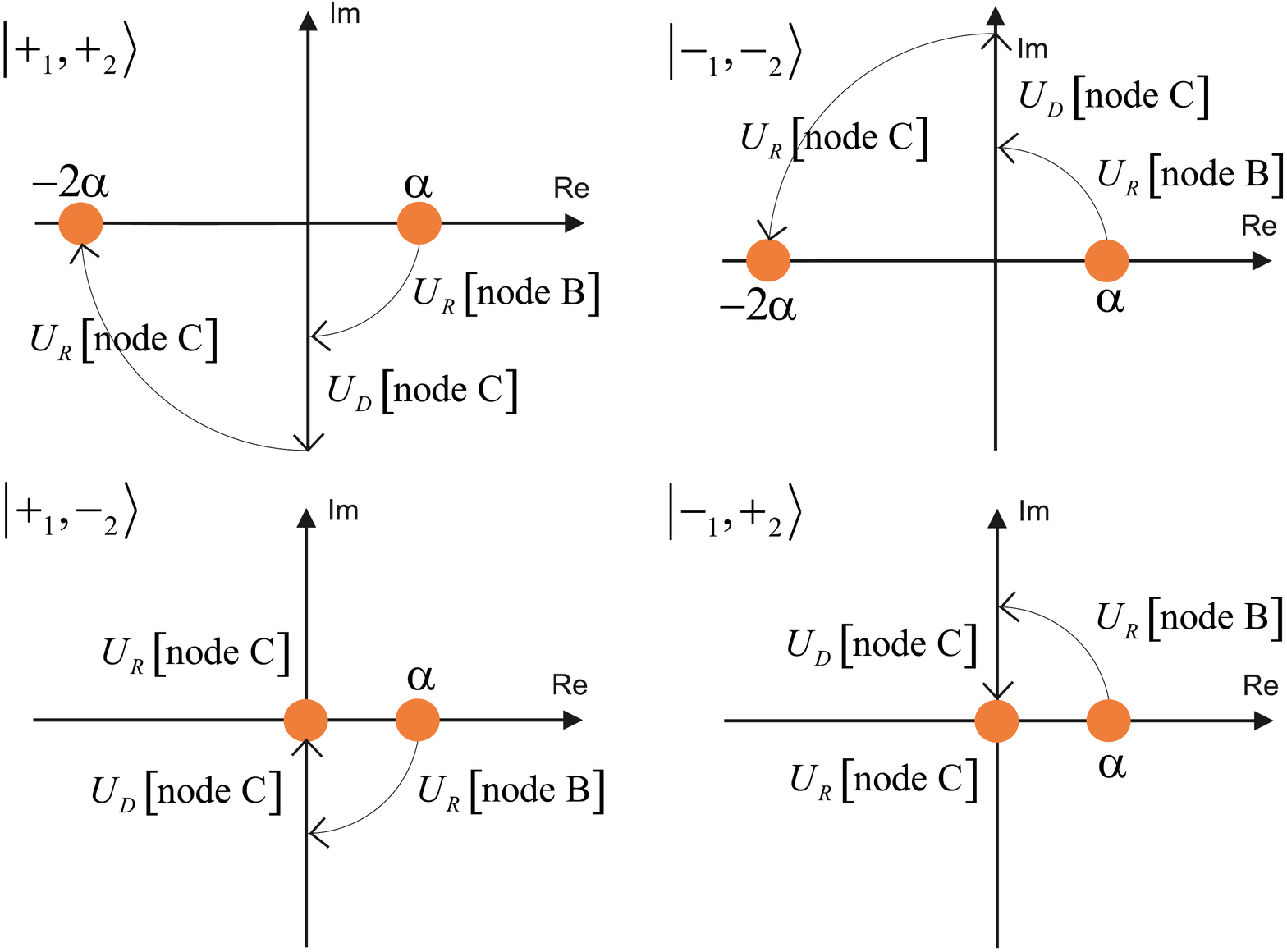} \\
\caption{(Color online) For each product state of two atoms (in the basis $\{ \ket{+}, \ket{-} \}$),
we display the evolution of the input coherent state $\ket{\alpha}$ due to the sequence of three
operations, namely, (i) controlled phase shift $U_R(-\pi/2)$ in node B, (ii) controlled displacement
$U_D(-\im \alpha)$ in node C, and (iii) controlled phase shift $U_R(-\pi/2)$ in node C. In this figure, 
for simplicity, we set $\eta = 1$ (lossless case).}
\label{fig5}
\end{center}
\end{figure}
\subsection{Entanglement Distribution}
\label{sec:2}

In this part of setup, atoms $1$ and $2$ are extracted from the MOTs of each repeater node 
and inserted into the S-conveyors. Both atoms are initialized in the ground state $\ket{0}$ 
and arrive simultaneously in the cavities $C_1$ and $C_4$. Each atom encodes a qubit by means 
of a three-level atom in the $\Lambda$ configuration as displayed in Fig.~\ref{fig1}(c). In 
order to protect this qubit against the decoherence, the (qubit) states $\ket{0}$ and $\ket{1}$ 
are chosen as the stable ground and long-living metastable atomic states or as the two hyperfine 
levels of the ground state.

Once conveyed into the cavity $C_1$ ($C_4$), atom $1$ ($2$) couples simultaneously to the 
photon field of cavity and two continuous laser beams as displayed in Fig.~\ref{fig1}(c). 
The evolution of the coupled atom-cavity-laser system in the node B is governed by
the Hamiltonian
\begin{equation}\label{ham-rot}
H_{R} = \frac{\hbar \, J_2}{2} \, \sigma^X a^\dagger a,
\end{equation}
where $\sigma^X$ is the respective Pauli operator in the computational basis $\{\ket{0}, 
\ket{1} \}$, while $J_2$ is the effective coupling (see Appendix, where we 
show that the above Hamiltonian is produced deterministically in our setup). The evolution 
governed by $H_{R}$ reads
\begin{equation}\label{evol-rot}
U_{R}(\theta) = e^{\im \theta \, \sigma^X a^\dagger a},
\end{equation}
where $\theta = - J_2 \, t / 2$. This evolution implies a phase shift of the cavity field 
by the angle $\theta$ or $(- \, \theta)$ conditioned upon the atomic state, where $\theta$
is proportional to the atom-cavity-laser evolution time that, in turn, is inversely proportional 
to the velocity of conveyed atom.

In contrast to node B, the evolution of the coupled atom-cavity-laser system in the node 
C is governed by a sequence of two evolutions. While the second evolution is governed by 
the Hamiltonian (\ref{ham-rot}), the first evolution is governed by the Hamiltonian
\begin{equation}\label{ham-displ}
H_{D} = \frac{\hbar \, J_1}{2} \left( a + a^\dagger \right) \sigma^X ,
\end{equation}
where $J_1$ is the atom-field coupling (see Appendix, where we show that the 
above Hamiltonian is produced deterministically in our setup). The evolution governed by 
$H_{D}$ reads
\begin{equation}\label{evol-displ}
U_{D}(\beta) = e^{\left( \beta \, a^\dag - \beta^* a \right) \, \sigma^X}
              = D \left( \beta \, \sigma^X \right) \, ,
\end{equation}
where $\beta = - \im J_1 \, t / 2$. This evolution implies a displacement of the cavity field 
mode by the amount $\beta$ conditioned upon the atomic state, where $\beta$ is proportional 
to the atom-cavity-laser evolution time that, in turn, is inversely proportional to the velocity 
of conveyed atom. We remark that both Hamiltonians imply that the (fast-decaying) excited state 
$\ket{e}$ remains almost unpopulated during the respective evolutions.

First, a pulse of a weak coherent light produced by the source $P_1$ and characterized 
by a real amplitude $\alpha$ interacts with the atom-cavity system in node B, where the 
cavity is prepared in the vacuum state. The evolution (\ref{evol-rot}) leads to the 
atom-light entangled state
\begin{equation}
U_{R} (\theta) \ket{\alpha} \ket{0_1} = 
	\frac{1}{\sqrt{2}} \left( \ket{e^{\im \theta} \alpha} \ket{+_1} 
	+ \ket{e^{-\im \theta} \alpha} \ket{-_1} \right),
\end{equation}
where $\ket{+} = \left( \ket{0} + \ket{1} \right) / \sqrt{2}$ and $\ket{-} = \left( 
\ket{0} - \ket{1} \right) / \sqrt{2}$. The resulting coherent state from the cavity is 
outputted into the transmission channel between the nodes. Since we are dealing with 
a high-finesse cavity and since the fast-decaying atomic state $\ket{e}$ remains almost 
unpopulated during the evolution, the dominant photon loss occurs in the optical fiber 
connecting the cavities $C_1$ and $C_2$ that plays the role of a transmission 
channel in our setup. Since photon loss increases with the length of the fiber, to a good 
approximation, we describe the loss using a beam-splitter model that transmits only 
a part of the coherent light pulse through the channel,
\begin{equation}\label{loss}
\ket{vac}_E \, \ket{\alpha} \longrightarrow
 \ket{\sqrt{1 - \eta} \, \alpha}_E \, \ket{\sqrt{\eta} \, \alpha} \, ,
\end{equation}
where the subscript $E$ refers to an environmental light mode responsible for 
the fiber relaxation. In this expression, $\eta = e^{-\ell / \ell_\circ}$ 
describes the attenuation of the transmitted light through the fiber, where 
$\ell$ is the distance between the repeater nodes, while $\ell_\circ$ 
is the attenuation length that can reach almost $25$ km for fused-silica fibers 
at telecommunication wavelengths. Below we set $\ell_\circ = 25$ km.

Next, the attenuated light pulse interacts with the atom-cavity system in node 
C, where (as in node B) the cavity is prepared in the vacuum state. By tracing 
over the environmental degrees of freedom (mode with the subscript $E$), the 
evolution $U_{R} (-\pi /2)$ in node B along with the sequence of evolutions $U_{D} 
(- \im \sqrt{\eta} \, \alpha)$ and $U_{R} (-\pi /2)$ in node C followed by an 
unconditional displacement $D(\alpha \, \sqrt{\eta})$ leads to the entangled state 
between the atoms and optical modes (see Fig.~\ref{fig5})
\begin{equation}\label{state0}
\rho = \frac{1 + e^{- 2 \, \alpha^2 (1 - \eta)}}{2} \ket{p} \bra{p} 
      + \frac{1 - e^{- 2 \, \alpha^2 (1 - \eta)}}{2} \ket{q} \bra{q} \, ,
\end{equation}
where
\begin{eqnarray}
\ket{p} = \frac{1}{\sqrt{2}} \left( \ket{\sqrt{\eta} \, \alpha} \ket{\phi_{1,2}^-}
            + \ket{- \sqrt{\eta} \, \alpha} \ket{\phi_{1,2}^+} \right) ;  \\
\ket{q} = \frac{1}{\sqrt{2}} \left( \ket{\sqrt{\eta} \, \alpha} \ket{\psi_{1,2}^-}
            - \ket{- \sqrt{\eta} \, \alpha} \ket{\psi_{1,2}^+} \right) ,        		
\end{eqnarray}
such that $\braket{p}{p} = \braket{q}{q} = 1$, while $\braket{p}{q} = 0$.

The (entangled with both atoms) light pulse is guided into the balanced beam 
splitter displayed in Fig.~(\ref{fig1})(b) and that is characterized by the 
transmissivity and reflectivity parameters $T = 1 / \sqrt{2}$ and $R = \im / \sqrt{2}$,
respectively. According to this coherent state discrimination (CSD) setup, the 
input pulse interferes with the (ancilla) coherent state $\ket{\im \sqrt{\eta} \, \alpha}$ 
sent into another input port of the beam splitter. It can be readily checked that the 
coherent states $\ket{\pm \sqrt{\eta} \, \alpha}$ encapsulated in Eq.~(\ref{state0}) and 
supplemented by the ancilla state are transformed due to the beam splitter as follows
\begin{equation}\label{transf}
\ket{\im \sqrt{\eta} \, \alpha} \ket{\sqrt{\eta} \, \alpha} 
				\rightarrow \ket{0} \ket{\im \sqrt{2 \, \eta} \, \alpha}, \quad
\ket{\im \sqrt{\eta} \, \alpha} \ket{- \sqrt{\eta} \, \alpha} 
				\rightarrow \ket{- \sqrt{2 \, \eta} \, \alpha} \ket{0} \, .
\end{equation}
On the right hand side of above equations, each ket vector in the expression corresponds to 
an output port of the beam splitter. Having two detectors resolving \textit{click} or 
\textit{no click} detection events, there are three possible detection patterns, since the 
pattern $\{$click, click$\}$ cannot happen. Among these patterns, the pattern $\{$no click, 
no click$\}$ is inconclusive, while the remaining two patterns $\{$click, no click$\}$ and 
$\{$no click, click$\}$ lead to the desired state discrimination.

We compute first the respective probabilities of success
\begin{eqnarray}\label{prob-succ1}
P^{dist} &=& 
\mbox{Tr} \left( \hat{\rho} \ket{\mbox{no click}, \mbox{click}} \bra{\mbox{no click}, 
	\mbox{click}} \right)\nonumber \\
&=& \mbox{Tr} \left( \hat{\rho} \ket{\mbox{click}, \mbox{no click}} \bra{\mbox{click}, 
	\mbox{no click}} \right) = \frac{1}{2} \left( 1 - e^{-2 \, \eta \, \alpha^2} \right) \, ,
\end{eqnarray}
where $\hat{\rho}$ denotes the state $\rho \, \otimes \ket{\im \sqrt{\eta} \, \alpha} 
\bra{\im \sqrt{\eta} \, \alpha} $ being transformed in concordance with Eq.~(\ref{transf}).
Conditioned upon above detection patterns, the density function $\hat{\rho}$ reduces to 
the one of two (atom-atom) entangled states
\begin{equation}\label{state1a}
\tilde{\rho}_{-,f}^{1,2} = \frac{\bra{\mbox{no click, click}} \, \hat{\rho} \, \ket{\mbox{no click, click}}}
{P^{dist}} = f \, \ket{\phi^-_{1,2}} \bra{\phi^-_{1,2}} + (1-f) \ket{\psi^-_{1,2}} \bra{\psi^-_{1,2}} \, ;
\end{equation}
or
\begin{equation}\label{state1b}
\tilde{\rho}_{+,f}^{1,2} = \frac{\bra{\mbox{click, no click}} \, \hat{\rho} \, \ket{\mbox{click, no click}}}
{P^{dist}} = f \, \ket{\phi^+_{1,2}} \bra{\phi^+_{1,2}} + (1-f) \ket{\psi^+_{1,2}} \bra{\psi^+_{1,2}} \, ,
\end{equation}
where the entanglement fidelity is given by the expression
\begin{equation}\label{fidelity}
f = \frac{1}{2} \left( 1 + e^{- 2 \, \alpha^2 (1 - \eta)} \right) \, .
\end{equation}
If the output of photon detection yields the inconclusive pattern $\{$no click, no click$\}$, 
then the atoms $1$ and $2$ should be discarded and the entanglement distribution protocol 
repeated using the next atomic pair conveyed by means of S-conveyors along the setup in both 
repeater nodes.

In this section, we developed a feasible scheme for distribution of atomic entanglement that 
is conditioned upon two detection patterns indicated above, where the total success probability 
associated with these patterns is $2 \, P^{dist}$, while the fidelity is given by the above 
expression. We exploit the controlled phase shift (\ref{evol-rot}) in both repeater nodes supplemented
by a controlled displacement in the second node. In contrast to Ref.~\cite{pra78} and our previous 
paper \cite{pra88a}, this combined approach greatly simplifies the tripartite entangled state 
of light and two atoms [compare Eq.~(\ref{state0}) in this paper to Eqs.~(14) and (4) in the 
above mentioned references]. More specifically, Eq.~(\ref{state0}) encapsulates only two optical 
coherent states, which differ by a minus sign and are, therefore, much easier to discriminate. 
This simplification, in turn, implies a rather uncomplicated discrimination setup (CSD) displayed 
in Fig.~\ref{fig1}(b) consisting of a single beam-splitter and two photon detectors to resolve 
click and no-click events.

We emphasize that in the above cited references, the entangled state of light and two atoms 
involve three different coherent states leading to a more demanding discrimination scheme. 
Specifically, the discrimination scheme in Ref.~\cite{pra78} requires two beam splitters 
and three unconditional displacements together with three photon detectors capable to resolve 
click and no click events. Although the respective CSD scheme in our previous paper \cite{pra88a} 
requires a single beam splitter and one unconditional displacement, a notable difficulty poses 
the requirement to use photon number resolving detectors. As a consequence of simplified CSD in 
this paper, we obtain a much higher success probability (compare Eq.~(\ref{prob-succ1}) in 
this paper to $\mathcal{N}_{-}/4$ in Ref.~\cite{pra88a} and to $P^{total,USD}$ and $P^{total,ent}$ 
in Ref.~\cite{pra78}) leading, in turn, to much higher repeater rates. We remark that, in 
contrast to the controlled phase shift used in Ref.~\cite{pra78}, in this paper, we exploit 
the evolution (\ref{evol-rot}) based on $\sigma^X$ instead of $\sigma^Z$ and, moreover, we 
avoid manipulations of atomic states associated with the produced entangled state.

Finally, we note that Eq.~(\ref{fidelity}) coincides with the respective fidelity 
obtained in our previous paper, in which we have mentioned that this fidelity is close to 
one when $\alpha^2 (1 - \eta) \ll 1$. Since we considered one single purification round, this 
restriction led us to the regime $\alpha^2 \leq 1$ that we considered through the paper. Due 
to four purification rounds considered in this paper, we succeed to relax this restriction and 
consider $\alpha^2 = 1,2,3$ (see below).

\subsection{Entanglement Purification}
\label{sec:3}

Assuming that the entanglement distribution was successful, atoms $1$ and $2$ are conveyed into 
the area, in which they are subjected to the (single-qubit) Hadamard transformation. Due to this 
unitary transformation, the states (\ref{state1a}) and (\ref{state1b}) take the form
\begin{eqnarray}
\rho_{-,f}^{1,2} &=& f \, \ket{\psi^+_{1,2}} \bra{\psi^+_{1,2}} + (1-f) \ket{\psi^-_{1,2}} \bra{\psi^-_{1,2}} \, , 
\label{state2a} \\
\rho_{+,f}^{1,2} &=& f \, \ket{\phi^+_{1,2}} \bra{\phi^+_{1,2}} + (1-f) \ket{\phi^-_{1,2}} \bra{\phi^-_{1,2}} \, . 
\label{state2b}
\end{eqnarray}
Once atoms $1$ and $2$ enter the cavities $C_2$ and $C_5$, their (conveyed) velocities 
decrease gradually until zero, such that atoms remain trapped right inside the respective cavities.

At the same time, atoms $3$ and $4$ are inserted from MOTs into V-conveyors of both 
repeater nodes and transported with a constant velocity along the setup. Similar to atoms 
$1$ and $2$, this atomic pair is entangled and subjected then to the Hadamard gate. 
Assuming that the second entanglement distribution was successful, atoms $3$ and $4$ are 
now described by the state $\rho_{-,f}^{3,4}$ or $\rho_{+,f}^{3,4}$ having the structure
of Eqs.~(\ref{state2a}) or (\ref{state2b}), respectively.
Atoms $3$ and $4$ are conveyed along the setup until they couple (simultaneously) to 
cavities $C_2$ and $C_5$ prepared both in the vacuum state. At this point, each of these 
cavities shares a pair of atoms $1$, $3$, and $2$, $4$, respectively.

Inside the cavities $C_2$ and $C_5$, these atomic pairs evolve due to the interaction 
governed by the Heisenberg XX Hamiltonians
\begin{equation}\label{ham-xx}
H^{1,3}_{XX} = \frac{\hbar \, J_3}{2} \, \sigma_1^X \sigma_3^X \, , \quad \mbox{and} \quad
H^{2,4}_{XX} = \frac{\hbar \, J_3}{2} \, \sigma_2^X \sigma_4^X \, ,
\end{equation}
over the time period $T = \pi / (2 \, J_3)$, where $J_3$ is the effective coupling. The 
resulting density functions
\begin{eqnarray}
\rho_{++,f}^{1-4} &=& e^{- \im H^{1,3}_{XX} \, T / \hbar} e^{- \im H^{2,4}_{XX} \, T / \hbar} \left( \rho_{+,f}^{1,2} 
				\otimes \rho_{+,f}^{3,4} \right) e^{\im H^{2,4}_{XX} \, T / \hbar} e^{\im H^{1,3}_{XX} \, T / \hbar} \, ; \quad \\
\rho_{--,f}^{1-4} &=& e^{- \im H^{1,3}_{XX} \, T / \hbar} e^{- \im H^{2,4}_{XX} \, T / \hbar} \left( \rho_{+,f}^{1,2}
				\otimes \rho_{-,f}^{3,4} \right) e^{\im H^{2,4}_{XX} \, T / \hbar} e^{\im H^{1,3}_{XX} \, T / \hbar} \, ; \quad \\
\rho_{+-,f}^{1-4} &=& e^{- \im H^{1,3}_{XX} \, T / \hbar} e^{- \im H^{2,4}_{XX} \, T / \hbar} \left( \rho_{+,f}^{1,2}
				\otimes \rho_{-,f}^{3,4} \right) e^{\im H^{2,4}_{XX} \, T / \hbar} e^{\im H^{1,3}_{XX} \, T / \hbar} \, ; \quad \\
\rho_{-+,f}^{1-4} &=& e^{- \im H^{1,3}_{XX} \, T / \hbar} e^{- \im H^{2,4}_{XX} \, T / \hbar} \left( \rho_{+,f}^{1,2}
				\otimes \rho_{+,f}^{3,4} \right) e^{\im H^{2,4}_{XX} \, T / \hbar} e^{\im H^{1,3}_{XX} \, T / \hbar} \, ,
\end{eqnarray}
describe four-partite entangled states.
We remark that the above Heisenberg XX Hamiltonian is produced deterministically in our 
setup by coupling atoms (simultaneously) to the same cavity mode and two laser beams in 
the strong driving regime (see Appendix A of Ref.~\cite{pra88a}).

\begin{figure}[!t]
\begin{center}
\includegraphics[width=0.95\textwidth]{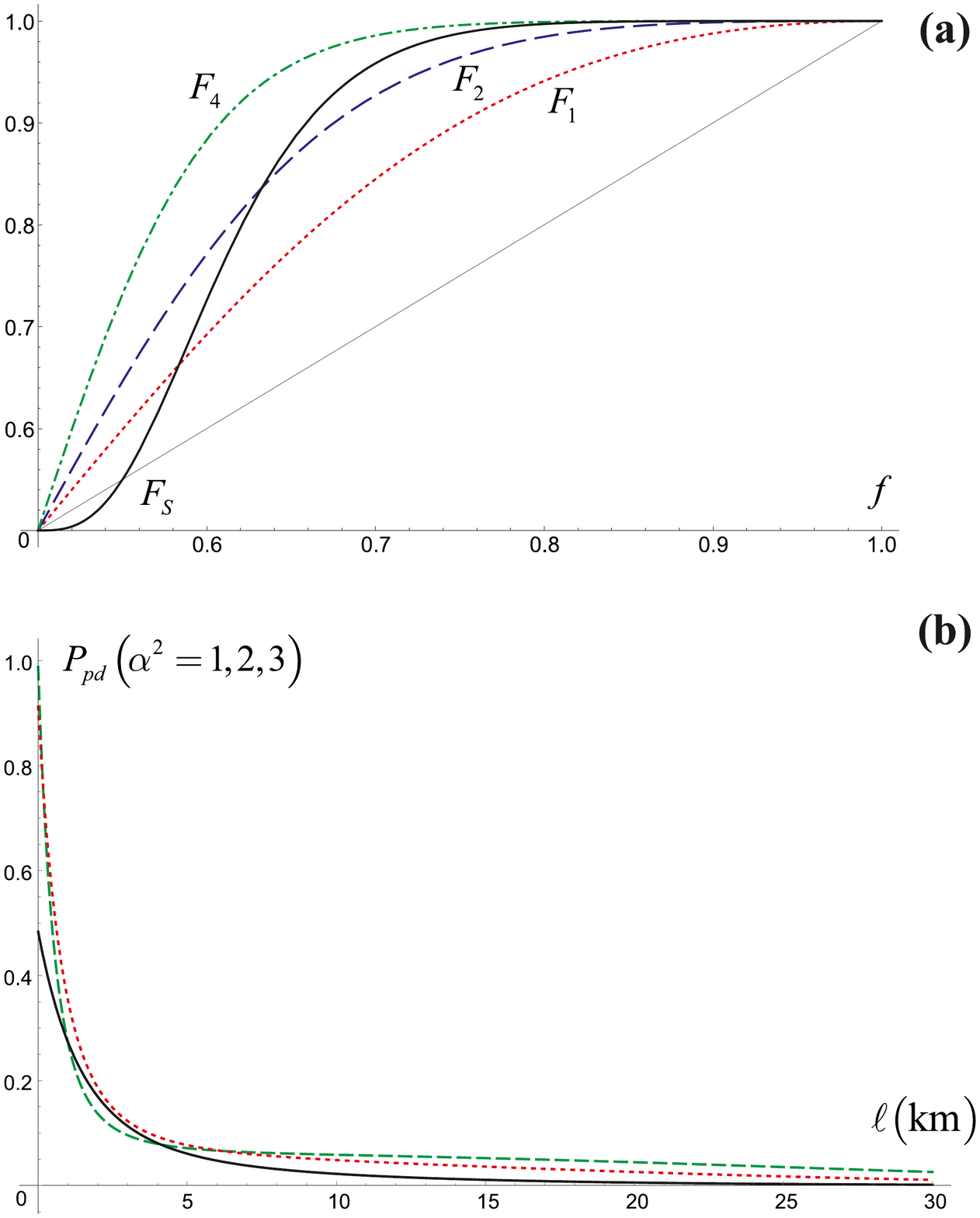} \\
\caption{(Color online) (a) Purified fidelities $F_1, F_2$, and $F_4$ along with the swapped 
fidelity $F_S$ as functions of the initial fidelity (\ref{fidelity}). (b) Success probability 
$P_{pd}$ associated with the entanglement distribution and purification as a function of elementary 
repeater distance $\ell$ displayed for $\alpha^2 = 1$ (solid curve), $\alpha^2 = 2$ (dotted curve), 
and $\alpha^2 = 3$ (dashed curve).}
\label{fig2}
\end{center}
\end{figure}

Being further conveyed along the setup, atoms $3$ and $4$ decouple from cavities $C_2$ 
and $C_5$, and get projected outside in the computational basis. Purification protocol 
is successful if the outcome of atomic projection reads
\begin{eqnarray}
\{ 0, 1 \} \quad \mbox{or} \quad \{ 1, 0 \} \quad \mbox{for} \quad \rho_{++,f}^{1-4} \quad
\mbox{and} \quad \rho_{-+,f}^{1-4} \, , \label{pattern2a} \\
\{ 0, 0 \} \quad \mbox{or} \quad \{ 1, 1 \} \quad \mbox{for} \quad \rho_{--,f}^{1-4} \quad
\mbox{and} \quad \rho_{+-,f}^{1-4} \, . \label{pattern2b}
\end{eqnarray}
In this case, above detection patterns lead to the reduced density functions
\begin{eqnarray}\label{state4}
\rho_{-,F_1}^{1,2} &=& \frac{\bra{0_3, 1_4} \rho_{++,f}^{1-4} \ket{0_3, 1_4}}{P^{purif}_1} 
			 = \frac{\bra{1_3, 0_4} \rho_{++,f}^{1-4} \ket{1_3, 0_4}}{P^{purif}_1}
			 = \frac{\bra{0_3, 0_4} \rho_{+-,f}^{1-4} \ket{0_3, 0_4}}{P^{purif}_1} \nonumber \\
			 &=& \frac{\bra{1_3, 1_4} \rho_{+-,f}^{1-4} \ket{1_3, 1_4}}{P^{purif}_1} 
			 = F_1 \ket{\psi^+_{1,2}} \bra{\psi^+_{1,2}} + (1 - F_1) \ket{\psi^-_{1,2}} 
			 \bra{\psi^-_{1,2}} \, ; \\
\rho_{+,F_1}^{1,2} &=& \frac{\bra{0_3, 1_4} \rho_{-+,f}^{1-4} \ket{0_3, 1_4}}{P^{purif}_1} 
			 = \frac{\bra{1_3, 0_4} \rho_{-+,f}^{1-4} \ket{1_3, 0_4}}{P^{purif}_1}
			 = \frac{\bra{0_3, 0_4} \rho_{--,f}^{1-4} \ket{0_3, 0_4}}{P^{purif}_1} \nonumber \\
			 &=& \frac{\bra{1_3, 1_4} \rho_{--,f}^{1-4} \ket{1_3, 1_4}}{P^{purif}_1} 
			 = F_1 \ket{\phi^+_{1,2}} \bra{\phi^+_{1,2}} + (1 - F_1) \ket{\phi^-_{1,2}} 
			 \bra{\phi^-_{1,2}} \, ,
\end{eqnarray}
where the purified fidelity and the success probability take the form
\begin{equation}\label{fidelity1}
F_1 = \frac{f^2}{1 + 2 \left( f^2 - f \right)} \, , \quad P^{purif}_1 = \frac{1}{2} - f + f^2 \, .
\end{equation}
The above purified states preserves the rank 2 form and are characterized 
by the fidelity (\ref{fidelity1}) displayed in Fig.~\ref{fig2}(a) by a dotted curve. We remark 
that if the outcome of atomic projection disagrees with the patterns (\ref{pattern2a}) and
(\ref{pattern2b}), the entanglement purification is considered unsuccessful. In this case, we 
discard all atoms and repeat both distribution and purification protocols using fresh atomic 
pairs conveyed from MOTs in each repeater node.

Assuming that the entanglement purification is successful, atoms $5$ and $6$ are inserted from 
MOTs into V-conveyors and transported with a constant velocity along the setup. Similar to atoms 
$3$ and $4$, this atomic pair gets entangled, subjected to the Hadamard gate, and conveyed 
further into the cavities $C_2$ and $C_5$, such that each of these cavities shares a pair of 
atoms $1$, $5$, and $2$, $6$, respectively. In these cavities, atoms evolve due to the 
Heisenberg XX interaction over the time period $T$. Being further conveyed along the setup, 
atoms $5$ and $6$ are then projected in the computational basis. As before, the entanglement 
purification is successful if the outcome of atomic projection coincides with (\ref{pattern2a})
and (\ref{pattern2b}). In this case, we get the reduced density functions
\begin{eqnarray}
\rho_{-,F_2}^{1,2} &=& F_2 \ket{\psi^+_{1,2}} \bra{\psi^+_{1,2}} + (1 - F_2) \ket{\psi^-_{1,2}} 
					\bra{\psi^-_{1,2}} \, , \\
\rho_{+,F_2}^{1,2} &=& F_2 \ket{\phi^+_{1,2}} \bra{\phi^+_{1,2}} + (1 - F_2) \ket{\phi^-_{1,2}} 
					\bra{\phi^-_{1,2}} \, ,
\end{eqnarray}
where the purified fidelity $F_2$ and the success probability $P^{purif}_2$ are computed 
using the iterative formulas
\begin{eqnarray}\label{formulas}
F_n &=& \frac{f \, F_{n-1}}{1 - F_{n-1} + f \left( 2 \, F_{n-1} - 1 \right)} \, , \label{formulas1} \\
P^{purif}_n &=& \frac{1}{2} 
				\left( 1 - F_{n-1} + f \left( 2 \, F_{n-1} - 1 \right) \right) \, , \label{formulas2}
\end{eqnarray}
where $F_0 \equiv f$, while $F_2$ is displayed in Fig.~\ref{fig2}(a) by a dashed curve.

Assuming that the third purification (using atoms $7$, $8$) is successful, atoms $9$, $10$ are 
conveyed along the setup and coupled then to the cavities $C_2$, $C_5$ in order to realize the last 
(fourth) purification round. In contrast to the sequence we described above, atoms $9$, $10$ are 
not projected after they leave the respective cavities. Instead, atoms $1$ and $2$ are projected 
in the computational basis using the (nondestructive) projective measurements directly inside the 
cavities (see Sec.~II.D below). In contrast to the patterns (\ref{pattern2a}) and (\ref{pattern2b}), 
the entanglement purification is successful if the outcome of atomic projection (atoms $1$ and $2$) 
reads 
\begin{eqnarray}
\{ 0, 1 \} \quad \mbox{or} \quad \{ 1, 0 \} \quad \mbox{for} \quad \rho_{++,F_3}^{1,2,9,10} \quad
\mbox{and} \quad \rho_{+-,F_3}^{1,2,9,10} \, , \label{pattern3a} \\
\{ 0, 0 \} \quad \mbox{or} \quad \{ 1, 1 \} \quad \mbox{for} \quad \rho_{--,F_3}^{1,2,9,10} \quad
\mbox{and} \quad \rho_{-+,F_3}^{1,2,9,10} \, . \label{pattern3b}
\end{eqnarray}
With the success probability $P^{purif}_4$, above patterns yield the reduced density functions 
\begin{eqnarray}
\rho_{-,F_4}^{9,10} &=& F_4 \ket{\psi^+_{9,10}} \bra{\psi^+_{9,10}} + (1 - F_4) \ket{\psi^-_{9,10}} 
					\bra{\psi^-_{9,10}} \, ; \label{state6a} \\
\rho_{+,F_4}^{9,10} &=& F_4 \ket{\phi^+_{9,10}} \bra{\phi^+_{9,10}} + (1 - F_4) \ket{\phi^-_{9,10}} 
					\bra{\phi^-_{9,10}} \, , \label{state6b}
\end{eqnarray}
completely characterized by the fidelity $F_4$ displayed in Fig.~\ref{fig2}(a) by a dot-dashed curve. 
It is readily seen that we obtain an almost-unit purified fidelity for $f \geq 0.75$ using four 
(successful) purification iterations.

In this section, we developed a high-fidelity scheme for entanglement purification that leads
(iteratively) to a gradual growth of the purified fidelity, and where the total probability 
of success
\begin{eqnarray}\label{prob-fin}
P_{pd} &=& 2^4 \, P^{purif}_1 \, P^{purif}_2 \, P^{purif}_3 \, P^{purif}_4
 	( 2 \, P^{dist})^5 \nonumber \\ 
	&=& \left( 15 \, f^2 - 180 \, f^3 + 1130 \, f^4 - 4700 \, f^5 + 14088 \, f^6 - 31584 \, f^7 \right. \nonumber \\
	&+& 53776 \, f^8 - 69600 \, f^9 + 67648 \, f^{10} - 48000 \, f^{11} + 23552 \, f^{12} \nonumber \\ 
	&-& \left. 7168 \, f^{13} + 1024 \, f^{14} \right) \left( 1 - e^{-2 \, \eta \, \alpha^2 } \right)^5 \, ,
\end{eqnarray}
is associated with four purification rounds which, in turn, encapsulates success probabilities of 
five entanglement distributions. By inserting Eq.~(\ref{fidelity}) in the above expression, we 
display $P_{pd}$ in Fig.~\ref{fig2}(b) as a function of $\ell$ (encoded by $\eta$) for $\alpha^2 = 1$, 
$2$, and $3$. As expected, the total success probability decreases with the growth of the distance 
between the nodes. We emphasize that the sequence of steps utilized in this section can be related to 
the entanglement purification scheme C (so-called entanglement pumping) suggested in Ref.~\cite{pra59}, 
in which the fidelity of a single low-fidelity entangled pair becomes gradually increased at the cost 
of all other low-fidelity entangled pairs available. Indeed, in our approach we also produced first 
the entangled pair (\ref{state2a}) or (\ref{state2b}) described by the fidelity $F_0$ that became 
gradually increased up to $F_4$ at the cost of entangled pairs associated with the atoms $3, \ldots, 
10$ and each described by the fidelity $F_0$ as well.

In each node of proposed purification protocol we exploit one evolution governed by the Heisenberg 
XX Hamiltonian (17) followed by a projection of a single atom in the computation basis. In contrast 
to our previous paper \cite{pra88a}, in which we have utilized (i) a four-partite entangled state 
[see Eq.~(8) in the above reference] generated in a probabilistic fashion through the cat state 
discrimination, (ii) evolution governed by the Heisenberg XY Hamiltonian, and (iii) projection of 
two atoms in each repeater node, the proposed approach greatly simplifies our protocol since we 
avoid any ancilla state and we postselect only the atomic state detection events associated with two 
atoms (and not four as in the previous paper). Due to these improvements, in this paper, we succeed 
to increase significantly the success probability $P_{pd}$ associated with the entire purification 
protocol [compare Eq.~(\ref{prob-fin}) with Eq.~(26) in our previous paper].

We emphasize that a high purified fidelity obtained in this paper is the consequence of multiple 
purification rounds realized as an inherent part of our repeater scheme in this paper. Although we 
considered just four such rounds, our setup is capable to perform an arbitrary number of rounds at 
the cost of reduced success probability. The setup displayed in Fig. 1(a) of our previous paper, in 
contrast, is by design limited to a single purification round, such that an extension to four such 
rounds would require three additional cavities and six additional atoms in each repeater node along 
with three additional communication channels. Using this extended setup in the framework of approach 
proposed in our previous paper, we have checked that the purified fidelity obtained using four successive 
purification rounds is slightly smaller as the fidelity $F_4$ obtained in this paper. However, since 
one successful purification round is conditioned upon the generation of the ancilla state (8) and 
atomic postselection (15), we obtained a much lower probability of success if compared with $P_{pd}$ 
derived above. This result leads, in turn, to a dramatically low success probability associated with 
the entire scheme and almost vanishing repeater rates.

\subsection{Entanglement Swapping}
\label{sec:4}

Assuming that the entanglement purification was successful, the (high-fidelity) entangled atoms 
$9$ and $10$ are conveyed along the setup to the swapping region displayed in the bottom 
rectangle of Fig.~\ref{fig1}(a). Here, atoms couple the cavities $C_3$ and $C_6$, both 
prepared in vacuum state. The conveyed atoms together with the trapped atoms form two atomic pairs 
$9$, $12$, and $10$, $13$. We recall that atoms $12$ and $13$ are entangled with atoms $11$ and 
$14$, respectively, where each pair is described by the mixed states $\rho^{11,12}_{\pm,F_4}$
and $\rho^{13,14}_{\pm,F_4}$ having the structure of Eqs.~(\ref{state6a}) and (\ref{state6b}).

According to the entanglement swapping protocol we proposed in Ref.~\cite{pra88a}, each atomic 
pair evolves in cavities $C_3$ and $C_6$ due to the Heisenberg XX interaction utilized in the
previous section. We remark that being applied on the separate states of an atomic pair (in 
the computational basis) over the time period $T = \pi / (2 \, J_2)$, this evolution implies
\begin{eqnarray}
\ket{1, 1} &\longrightarrow& \frac{- \im}{\sqrt{2}} \left( \ket{0, 0} + \im \ket{1, 1} \right) \, ; \\
\ket{0, 0} &\longrightarrow& \frac{1}{\sqrt{2}} \left( \ket{0, 0} - \im \ket{1, 1} \right) \, ; \\
\ket{1, 0} &\longrightarrow& \frac{- \im}{\sqrt{2}} \left( \ket{0, 1} + \im \ket{1, 0} \right) \, ; \\
\ket{0, 1} &\longrightarrow& \frac{1}{\sqrt{2}} \left( \ket{0, 1} - \im \ket{1, 0} \right) \, , \quad
\end{eqnarray}
such that the resulting states form the modified Bell basis
\begin{equation}\label{mbasis}
 \ket{\mathbf{m}_{a,b}^k} = e^{- \frac{\im}{\hbar} H^{a,b}_{XX} \, T}
              \{ \ket{1_a, 1_b}, \ket{0_a, 0_b}, \ket{1_a, 0_b}, \ket{0_a, 1_b} \} \, , \nonumber
\end{equation}
where $k = 1,2,3,4$.

\begin{table}
\caption{Pairs of Bell states identified with $\varphi^{i,j}$ and $\chi^{i,j}$ for given 
$i$ (column) and $j$ (row).}
\label{tab1} 
\begin{tabular}{c | c c c c}
\hline
& 1 & 2 & 3 & 4 \\
\hline\noalign{\smallskip}
1 & $\psi^+ , \psi^-$ \, & $\psi^- , \psi^+$ \, & $\phi^+ , \phi^-$ \, & $\phi^- , \phi^+$ \\
2 & $\psi^- , \psi^+$ & $\psi^+ , \psi^-$ & $\phi^- , \phi^+$ & $\phi^+ , \phi^-$ \\
3 & $\phi^- , \phi^+$ & $\phi^+ , \phi^-$ & $\psi^- , \psi^+$ & $\psi^+ , \psi^-$ \\
4 & $\phi^+ , \phi^-$ & $\phi^- , \phi^+$ & $\psi^+ , \psi^-$ & $\psi^- , \psi^+$ \\
\noalign{\smallskip}\hline
\end{tabular}
\end{table}

This observation suggests a deterministic realization of entanglement swapping using just two 
consecutive steps (i) atomic pairs $9$, $12$, and $10$, $13$ are subjected to the evolutions 
$e^{- \frac{\im}{\hbar} H^{9,12}_{XX} \, T}$ and $e^{- \frac{\im}{\hbar} H^{10,13}_{XX} \, T}$, 
respectively, that is followed by (ii) projection of each atomic pair in the computational 
basis. Obviously, these two steps are equivalent with the atomic projection in the modified 
Bell basis (\ref{mbasis}), where the swapped state is given by the expression
\begin{eqnarray}\label{state6}
&& \frac{1}{\mathcal{N}}
            \bra{\mathbf{m}_{9,12}^i, \mathbf{m}_{10,13}^j}
            \left( \rho^{9,10}_{+,F_4} \otimes \rho^{11,12}_{+,F_4} \otimes \rho^{13,14}_{+,F_4} 
            \right) \ket{\mathbf{m}_{9,12}^i, \mathbf{m}_{10,13}^j} \nonumber \\
              && \quad = F_S \, \ket{\varphi^{i,j}_{11,14}} \bra{\varphi^{i,j}_{11,14}}
               + (1 - F_S) \ket{\chi^{i,j}_{11,14}} \bra{\chi^{i,j}_{11,14}} \, ,
\end{eqnarray}
taken here, for simplicity, for a particular case when all three entangled pairs have the 
structure of Eq.~(\ref{state6b}). In this expression, $\mathcal{N}$ is the normalization factor, 
the states $\varphi^{i,j}$ and $\chi^{i,j}$ are displayed in Table~\ref{tab1}, while the swapped 
fidelity takes the form
\begin{equation}\label{final}
F_S = F_4 (3 - 6 \, F_4 + 4 \, F^2_4) \, .
\end{equation}
As expected, the swapped density function (\ref{state6}) is diagonal in the standard Bell basis 
and is completely characterized by $F_S$ displayed in Fig.~\ref{fig2}(a) by a solid curve.

In this section, we employed the swapping protocol introduced in our previous paper. This 
protocol is deterministic and exploits the same evolution as utilized in the previous section 
along with atomic projections. In contrast to the entanglement distribution and purification 
protocols, however, the swapping protocol is entirely characterized by the success probability 
$P_{sw}$ that is affected mainly by the efficiency of atomic projective measurements and can be
close to one (see below). We conclude, therefore, that the total probability of success associated 
with the entanglement distribution, purification, and two swappings is given by the expression
\begin{equation}\label{prob}
P_{total} = P^2_{sw} \, P_{pd}  \, .
\end{equation}
\begin{figure}[!ht]
\begin{center}
\includegraphics[width=0.95\textwidth]{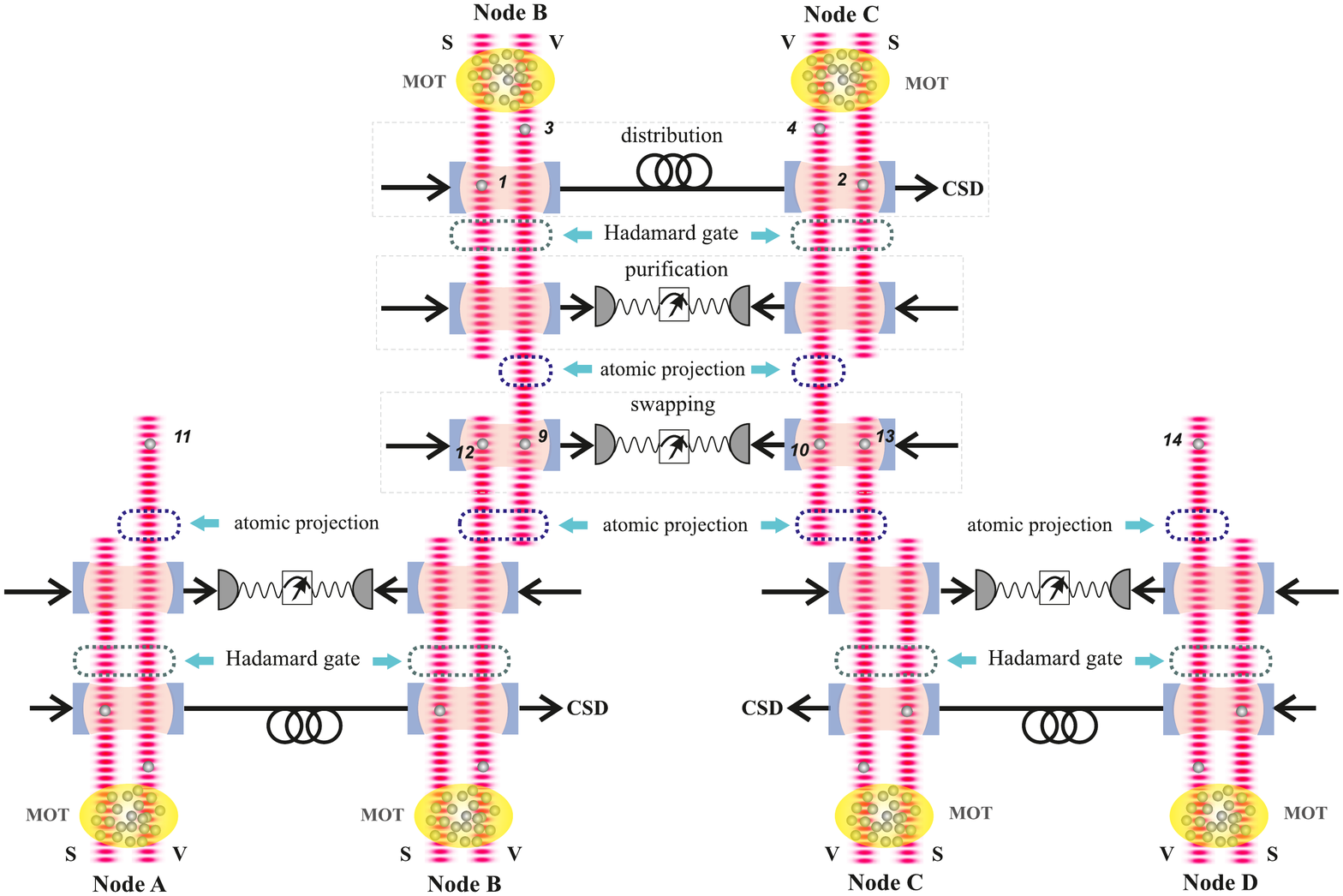} \\
\caption{(Color online) Sketch of experimental setup that realizes the
four-node repeater scheme.}
\label{fig3}
\end{center}
\end{figure}
\subsection{Remarks on the implementation of our scheme}
\label{sec:5}

For simplicity, in the setup displayed in Fig.~\ref{fig1}(a), we considered just two repeater 
nodes (B and C), where the atomic pairs $11$, $12$ and $13$, $14$ were initially entangled and 
given both by Eq.~(\ref{state4}). We are ready now to introduce the experimental setup that 
includes explicitly nodes A, B, C, and D. This setup is displayed in Fig.~\ref{fig3} and, in 
contrast to Fig.~\ref{fig1}(a), includes entanglement distribution and purification protocols 
associated with the (initially disentangled) atomic pairs $11$, $12$ and $13$, $14$. Simultaneously 
with the atomic pair $9$, $10$, these pairs are conveyed along the setup (but in the opposite 
direction) and follow the same sequence of interactions and projective measurements.

We recall that all three building blocks of our repeater require projective measurements of 
atoms which are located inside and outside the cavity. The method of atomic non-destructive 
measurements demonstrated in Refs.~\cite{prl97}, \cite{prl103} enables projective measurements of 
atoms coupled (strongly) to a cavity field that fits perfectly in our experimental setup.  
The physical mechanism behind these measurements exploits the suppression of cavity 
transmission due to the strong atom-cavity coupling. Recall that each atom in our scheme is 
a three-level atom in the $\Lambda$-configuration [see Fig.~\ref{fig1}(c)], where only the 
states $\ket{0}$ and $\ket{e}$ are coupled to the cavity field. If one such atom couples the 
cavity and is prepared in the $\ket{0}$ state, such that the cavity resonance is sufficiently 
detuned from the atomic $\ket{0} \leftrightarrow \ket{e}$ transition frequency, then the cavity
transmission drops according to the atom-cavity detuning and atom-cavity coupling. On the other 
hand, the cavity transmission remains unaffected if the atom was prepared in the state $\ket{1}$.

Once sufficiently many readouts of cavity transmission are recorded, this technique enables 
us to determine the state of a single atom with a high efficiency \cite{prl97}. 
Since the atom-cavity coupling increases proportionally with the number of loaded (into cavity) 
atoms, the same technique enables us to distinguish the following three separate states of two 
atoms (i) $\ket{0, 0}$, (ii) $\ket{0, 1}$ or $\ket{1, 0}$, and (iii) $\ket{1, 1}$ (see 
\cite{prl103}). We remark, however, that this approach cannot distinguish between the states 
$\ket{0, 1}$ and $\ket{1, 0}$ leading to an incompleteness of information about the swapped 
state (\ref{state6}) [see Table~\ref{tab1}]. In order to avoid this problem, one of atoms 
in $C_3$ ($C_6$) has to leave the cavity right after the inconclusive event occurs and be again 
projected outside the cavity.

The atomic projection outside the cavity is realized using a laser beam and a CCD camera 
located inside the oval regions displayed in Fig.~\ref{fig1}(a) below the cavities $C_2$ and 
$C_5$. While the laser beam removes atoms in a given quantum state from the conveyor without 
affecting atoms in the other state (so-called push-out technique \cite{prl91}), the CCD camera 
is used to detect the presence of remaining atoms via fluorescence imaging and determine, 
therefore, the state of each atom in question. In contrast to the previously discussed atomic 
projection, this technique is destructive and, thus, the projected atoms cannot be further
utilized.

\begin{figure}[!t]
\begin{center}
\includegraphics[width=0.95\textwidth]{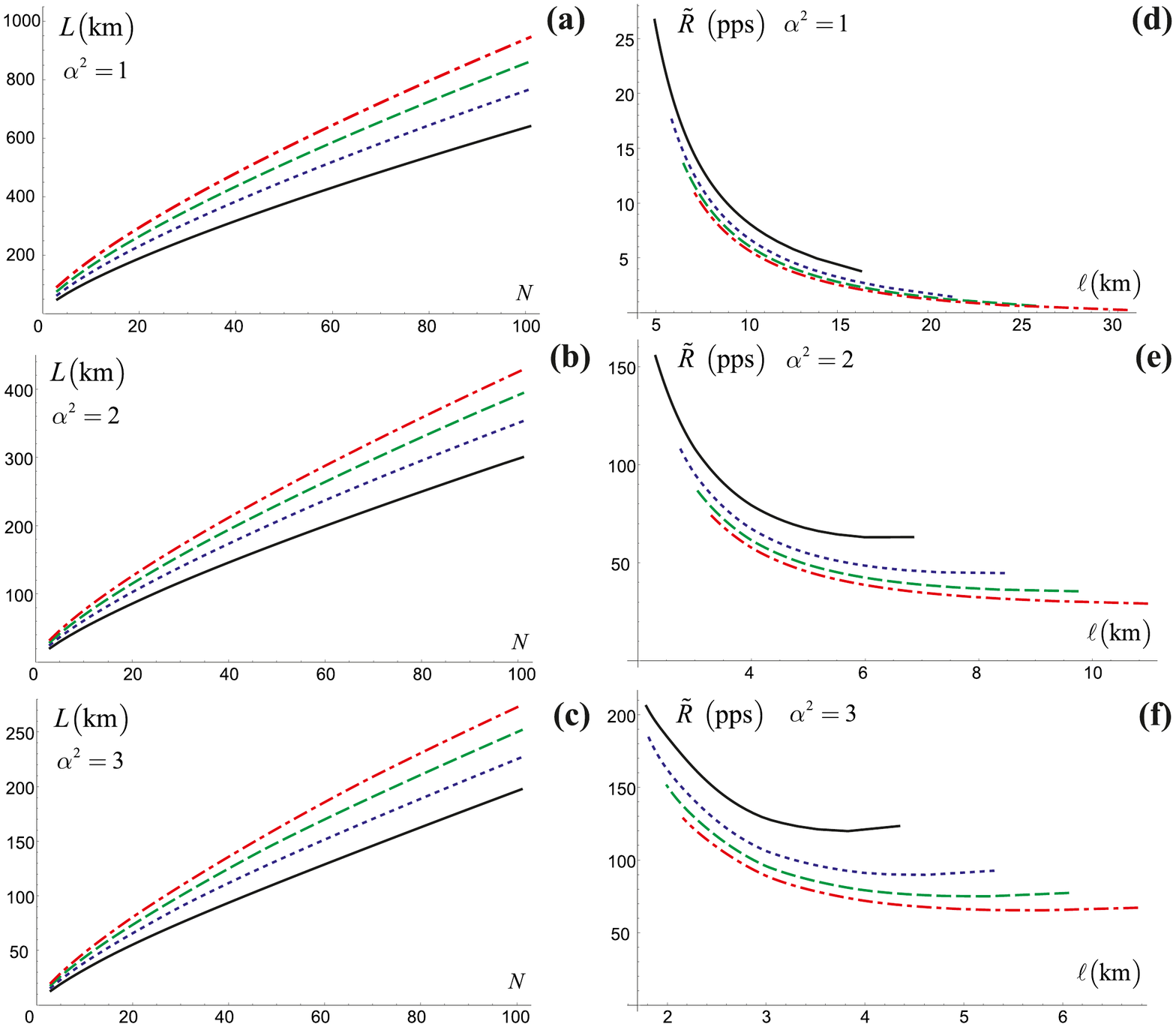} \\
\caption{(Color online) For the average photon numbers $\alpha^2 = 1$ (a), $2$ (b), and $3$
(c), we display the overall repeater distance $L = N \, \ell$ as a function of $N$ (number of 
segments) plotted for the final fidelities $F_{final} = 0.95$ (solid curve), $0.9$ (dotted curve), 
$0.85$ (dashed curve), and $0.8$ (dot-dashed curve). For the average photon numbers $\alpha^2 = 
1$ (d), $2$ (e), and $3$ (f), we display re-scaled repeater rates (\ref{rate2}) as functions
of $\ell$ (elementary length) plotted for the final fidelities $F_{final} = 0.95$ (solid curve), 
$0.9$ (dotted curve), $0.85$ (dashed curve), and $0.8$ (dot-dashed curve). We remark using plots 
(d)-(f) along with plots (a)-(c), one can easily calculate re-scaled repeater rates as functions 
of entire distance $L$.} 
\label{fig4}
\end{center}
\end{figure}
\section{Summary and Outlook}
\label{sec:6}

In the previous section, we have introduced our repeater scheme encapsulating three 
segments (four nodes) that corresponds to the overall distance of $3 \, \ell$. The 
extension of the sketch shown in Fig.~\ref{fig3} to an arbitrary number $N$ of 
segments is straightforward. With no loss of generality, we consider odd values of 
$N$ corresponding to $N+1$ repeater nodes or $N-1$ swappings. Setting $\alpha^2 = 1,
2$, and $3$, we display in Figs.~\ref{fig4}(a), (b), and (c) the overall distance as a 
function of $N$ taken for the final fidelities $F_{final} = 0.8, 0.85, 0.9, 0.95$, which 
are identified with the fidelity obtained after $N-1$ swapping operations. We infer from 
this figure that smaller values of $\alpha^2$ lead to larger overall distances, while the
segment length $\ell$ decreases with the increasing of $N$. We remark that this drawback 
is originated to the lack of repurification in our scheme that is supposed to act after 
each (or a few) swapping operation(s). We also emphasize that the obtained final fidelity 
is higher by about one order of magnitude as the respective fidelity obtained in our 
previous paper (compare, for instance, Fig.~\ref{fig4}(a) to Fig.~5(a) in 
Ref.~ \cite{pra88a}).

Besides final fidelities, we calculate the repeater rates which, together with the 
success probabilities displayed in Fig.~\ref{fig2}(b), provide the main set of 
characteristics associated with a quantum repeater. Since the atomic (fast-decaying) 
excited states remain almost unpopulated and each atomic qubit is encoded by other 
two (long-living) states, we assume that the atomic coherence times exceeds the overall 
time required to complete entanglement distribution, purification, and swapping in 
all repeater nodes. This assumption corresponds to a repeater with an ideal memory 
and implies that the only source of decoherence is the photon loss in the transmission 
channel.

We compute repeater rates (in units of pairs per second) using the following 
expression \cite{pra83} 
\begin{equation}\label{rate1}
R = \frac{1}{T_\circ \, Z_N (P_{total})} \, ,
\end{equation}
where $T_\circ$ is the total time required to distribute and purify an entangled 
pair over one repeater segment followed by two swappings, $N$ is the number of 
segments, while 
\begin{equation}
Z_N (P) = \sum_{j=1}^{N} \left( \begin{array}{c} N \\ j \end{array} \right)
		  \frac{(-1)^{j+1}}{1 - (1 - P)^j} \, . \nonumber
\end{equation}

According to the experimental setup displayed in Fig.~\ref{fig1}(a), the total 
time is composed of (i) interaction times required for entanglement distribution, 
purification, and swapping protocols, (ii) time required for both destructive and 
non-destructive atomic projections, and (iii) time required to communicate between 
the repeater nodes by means of coherent-state light and classical communication.
In contrast to the cases (i) and (iii), the time (iii) is readily computed using 
the relation $14 \, \ell / \upsilon$, where $\upsilon = 2 \times 10^8$ m/s is the 
velocity of light in the optical fiber. Without loss of generality, therefore, we
express the total time as $T_\circ = S \, 14 \, \ell / \upsilon$, where $S > 1$
is a real number encoding the times (i) and (iii) in terms of (iii). By inserting 
this expression into Eq.~(\ref{rate1}), we consider the re-scaled repeater rate
\begin{equation}\label{rate2}
R \, S = \frac{\upsilon}{14 \, \ell \, Z_N (P_{total})} \equiv \widetilde{R} \, .
\end{equation}

This re-scaled repeater rate is determined by the triplet $\{ \alpha, \ell, N \}$, 
which we extract from Fig.~\ref{fig4} for a given value of $F_{final}$. Setting 
$\alpha^2 = 1, 2$, and $3$, we display in Figs.~\ref{fig4}(d), (e), and (f) the 
re-scaled repeater rates as functions of $\ell$ taken for the final fidelities 
$F_{final} = 0.8, 0.85, 0.9$, and $0.95$. We infer from this figure that smaller 
values of $\alpha^2$ lead to smaller repeater rates $\widetilde{R}$. We remark that 
this behavior was expected since small $\alpha^2$ imply a larger total distance $L$, 
which in turn, leads to a reduction of the amount of produced entangled pairs (per 
second).

In this paper, a complete quantum repeater scheme that encapsulates entanglement 
distribution, purification, and swapping protocols was proposed. In contrast to 
conventional repeater schemes, we completely avoid two-qubit logical gates by 
exploiting cavity QED evolution along atomic projective measurements. Our repeater 
scheme has a conveyor like structure, 
in which single atoms are inserted into one of two optical lattices and conveyed along 
the entire repeater node. At the same time, another chain of atoms is conveyed along 
the neighboring repeater node in a synchronous fashion. These two nodes form together 
a repeater segment, while the entire set of segments form the quantum repeater itself.
In Figs.~\ref{fig1}(a) and \ref{fig3}, the sketch of experimental setup was displayed 
and a detailed description of all necessary steps and manipulations were provided. 
A comprehensive analysis of success probabilities and final fidelities obtained after
multiple purification and swapping operations was given. In addition, the comparison 
with regard to the results obtained in our previous paper has been provided.

In particular, our entanglement distribution here is based on a simpler coherent-state 
discrimination measurement distinguishing only two coherent states with opposite signs,
which can be easily and optimally done by linear optics and on-off detections. 
Following recent developments in cavity QED, finally, we briefly pointed to and 
discussed a few practical issues related to the implementation of our repeater scheme. 
We stress that although the physical resources utilized in our repeater are experimentally 
feasible, its explicit realization for a long-distance quantum communication still 
poses a serious challenge.

\vspace{2cm}
\textbf{Acknowledgments} \\ 

\noindent We thank the BMBF for support through the Q.Com (former QuOReP) program.

\appendix

\section{Derivation of Hamiltonians (\ref{ham-rot}) and (\ref{ham-displ})}
\label{app1}

In this appendix, we show that the evolutions (\ref{evol-rot}) and (\ref{evol-displ}) governed 
by the Hamiltonians (\ref{ham-rot}) and (\ref{ham-displ}) are produced deterministically in our 
framework. Namely, we consider an atom coupled strongly to the cavity and subjected to the detuned 
laser fields as displayed in Fig.~\ref{fig1}(c). The evolution of this interacting system is 
governed by the Hamiltonian 
\begin{eqnarray}\label{ham1}
H_1 &=& \hbar \, \omega_C \, a^\dag \, a - \im \hbar 
       \left[ \frac{g}{2} \, a \, \ket{e} \bra{0} 
       + \frac{\Omega}{2} \left(
       e^{-i \omega_L \, t} \ket{e} \bra{1} +
       e^{-i \omega_P \, t} \ket{e} \bra{0}\right) - H.c. \right] \nonumber \\
       &+& \hbar \left(
       \omega_1 \ket{1} \bra{1} +
       \omega_E \ket{e} \bra{e} +
       \omega_0 \ket{0} \bra{0} \right) \, ,
\end{eqnarray}
where $g$ denotes the coupling strength of atoms to the cavity mode, while $\Omega$ denotes 
the coupling strengths of atoms to both laser fields.

We switch to the interaction picture using the unitary transformation
\begin{equation}\label{picture1}
U_1 =    e^{- \im t \left[\left(
         \omega_1 \ket{1} \bra{1} + \left( \omega_1 + \omega_L + \Delta_L \right)
         \ket{e} \bra{e} + \omega_0 \ket{0} \bra{0} \right)
         + \left( \omega_1 + \omega_L - \omega \right) \, a^\dag a \right]}. \nonumber
\end{equation}
In this picture, the Hamiltonian (\ref{ham1}) takes the form
\begin{equation}\label{ham2}
H_2 = \hbar \, \Delta \, a^\dag a - \im \hbar 
       \left( \frac{g}{2} \, e^{-i \Delta_L \, t} a \, \ket{e} \bra{0}
    + \frac{\Omega}{2} \, e^{-i \Delta_L \, t} \left(
       \ket{e} \bra{1} + \ket{e} \bra{0} \right) - H.c. \right) .
\end{equation}
where the notation $\Delta_L \equiv (\omega_E - \omega_1) - \omega_L$, $\Delta_C \equiv 
(\omega_E - \omega_0) - \omega_C$, and $\Delta \equiv \Delta_L - \Delta_C$ has been 
introduced.

We require that $\Delta_L$ and $\Delta_C$ are sufficiently far detuned,
such that no atomic $\ket{e} \leftrightarrow \ket{0}$ or $\ket{e} \leftrightarrow \ket{1}$
transitions can occur. We expand the evolution governed by the
Hamiltonian (\ref{ham2}) in series and keep the terms up to the second order, that is,
\begin{equation}
U_2 \cong \vecI - \frac{\im}{\hbar} \int_{0}^{t} H_2 \, dt^\prime -
                    \frac{1}{\hbar^2} \int_{0}^{t} \left( H_2 \, \int_{0}^{t^\prime}
                    H_2 \, dt^{\prime \prime} \right) dt^\prime \, . \nonumber
\end{equation}
By performing integration and retaining only linear-in-time contributions,
we express this evolution in the form
\begin{equation}\label{operator2}
U_2 \cong \vecI - \frac{\im}{\hbar} \, H_3 \, t
    \cong \exp \left[ - \frac{\im}{\hbar} \, H_3 \, t \right],
\end{equation}
where the effective Hamiltonian is given by
\begin{equation}
H_3 = \hbar \, \Delta \, a^\dag a + \frac{\hbar \, \Omega}{4 \Delta_L} \left(
      \Omega \, \ket{1} \bra{0} + g \, \ket{1} \bra{0} \, a + H.c. \right) \nonumber
\end{equation}
after removing constant contributions. We switch to the interaction picture with respect 
to the first term of $H_3$. In this picture, we obtain
\begin{equation}\label{ham4}
H_4 = \frac{\hbar \, \Omega}{4 \Delta_L} \left(
      \Omega \, \ket{1} \bra{0} + g \, e^{-i \Delta \, t} \ket{1} \bra{0} \, a + H.c. \right) \, .
\end{equation}

We switch now from the atomic basis $\{ \ket{0}, \ket{1} \}$ to the basis
$\{ \ket{+}, \ket{-} \}$, where
\begin{equation}\label{basis}
\ket{+} = \frac{1}{\sqrt{2}} \left( \ket{0} + \ket{1} \right);  \quad
\ket{-} = \frac{1}{\sqrt{2}} \left( \ket{0} - \ket{1} \right).
\end{equation}
In this basis, the Hamiltonian (\ref{ham4}) takes the form
\begin{eqnarray}\label{ham5}
H_5 &=& \frac{\hbar \, \Omega}{8 \Delta_L} \left[ 2 \, \Omega S^Z
         + g \, S^Z ( e^{-i \Delta \, t} a + e^{i \Delta \, t} a^\dag ) \right. \nonumber \\
    && \hspace{1.5cm} \left. + \, g \, (S^\dag - S)(e^{-i \Delta \, t} a - e^{i \Delta \, t} 
       a^\dag) \right] , \qquad
\end{eqnarray}
where $S \equiv \ket{-} \bra{+}$ and $S^Z \equiv \ket{+} \bra{+} - \ket{-} \bra{-}$, and 
where we removed all the constant contributions. We switch one more time to the interaction 
picture with respect to the first term of (\ref{ham5}). In this picture, the Hamiltonian 
(\ref{ham5}) becomes
\begin{eqnarray}\label{ham6}
H_6 &=& \hbar \, \frac{g \, \Omega}{8 \Delta_L} 
        \left[ S^Z ( e^{-i \Delta \, t} a + e^{i \Delta \, t} a^\dag ) \right. \\
    && \left. + (S^\dag \, e^{\im \frac{\Omega^2}{2 \Delta_L} t}
    - S \, e^{-\im \frac{\Omega^2}{2 \Delta_L} t})(e^{-i \Delta \, t} a - e^{i \Delta \, t} 
    a^\dag) \right] . \nonumber
\end{eqnarray}

In the strong driving regime, i.e., for $\Omega^2 / (2\, \Delta_L) \gg \{ \Delta, \,
g \, \Omega / (8 \Delta_L) \}$, we eliminate the last (fast oscillating) term using the 
same arguments as for the rotating wave approximation. Using the identity $S^Z = \sigma^X$, 
the Hamiltonian (\ref{ham6}) reduces to
\begin{equation}\label{ham7}
H_7 = \hbar \, \frac{g \, \Omega}{8 \Delta_L}
      \left( e^{-i \Delta \, t} a + e^{i \Delta \, t} a^\dag \right) \sigma^X \, .
\end{equation}
In the case of vanishing $\Delta$ (equivalently $\Delta_L = \Delta_C$), the above
Hamiltonian takes the simplified form
\begin{equation}\label{ham8}
H_8 = \hbar \, \frac{g \, \Omega}{8 \Delta_L} \left(a + a^\dag \right) \sigma^X \, ,
\end{equation}
which, under the notation $J_1 \equiv g \, \Omega / (4 \, \Delta_L)$, coincides with
the Hamiltonian (\ref{ham-displ}).

In the case of non-vanishing $\Delta$, we introduce $\delta = \Delta - \Omega^2 / (2 \, 
\Delta_L)$, such that $\Omega^2 / (2 \Delta_L) > \delta$. Due to this assumption, we 
eliminate in Eq.~(\ref{ham6}) the first term along with the terms $S^\dag a^\dag e^{\im 
(\delta + \Omega^2 / \Delta_L ) t}$ and $S \, a \, e^{-\im (\delta + \Omega^2 / \Delta_L ) t}$ 
due to the rotating wave approximation, by which these fast oscillating terms play a minor 
role in the evolution. Without these terms, the Hamiltonian (\ref{ham6}) takes the usual 
Jaynes-Cummings form
\begin{equation}\label{ham9}
H_9 = \hbar \, \frac{g \, \Omega}{8 \Delta_L} \left( S^\dag a \, e^{-\im \delta \, t}  
		+ S \, a^\dag e^{\im \delta \, t} \right)  \, ,
\end{equation}
expressed in the (effective) atomic basis $\{ \ket{-}, \ket{+} \}$.

We require, finally, that $\delta$ is sufficiently far detuned, such that the transition 
$\ket{-} \leftrightarrow \ket{+}$ cannot occur. As above, we expand again the evolution 
governed by (\ref{ham9}) in series up to the second order and retain only linear-in-time 
contributions after the integration. This procedure leads to the effective Hamiltonian, 
\begin{equation}\label{ham10}
H_{10} = \hbar \, \frac{g^2 \, \Omega^2}{64 \, \Delta^2_L \, \delta} \left( \sigma^X a^\dag a 
			+ \sigma^X / 2 \right) \, ,
\end{equation}
where we removed all constant contributions and used the identity $S^Z = \sigma^X$. Since 
the second term commutes with the first term, we eliminate the second term by means of an 
appropriate interaction picture. The resulting Hamiltonian [that is, the first term of 
(\ref{ham10})], coincides with (\ref{ham-rot}) under the notation $J_2 = g^2 \, \Omega^2 / 
(32 \, \Delta^2_L \, \delta)$.

\end{document}